# AN EMPIRICAL EVALUATION OF IMPACT OF REFACTORING ON INTERNAL AND EXTERNAL MEASURES OF CODE QUALITY

S.H. Kannangara[1] and W.M.J.I. Wijayanake[2]

[1]School of Computing, National School of Business Management, Sri Lanka
[2]Department of Industrial Management, University of Kelaniya, Sri Lanka

## ABSTRACT

*Refactoring is the process of improving the design of existing code by changing its internal structure without affecting its external behaviour, with the main aims of improving the quality of software product. Therefore, there is a belief that refactoring improves quality factors such as understandability, flexibility, and reusability. However, there is limited empirical evidence to support such assumptions.*

*The objective of this study is to validate/invalidate the claims that refactoring improves software quality. The impact of selected refactoring techniques was assessed using both external and internal measures. Ten refactoring techniques were evaluated through experiments to assess external measures: Resource Utilization, Time Behaviour, Changeability and Analysability which are ISO external quality factors and five internal measures: Maintainability Index, Cyclomatic Complexity, Depth of Inheritance, Class Coupling and Lines of Code.*

*The result of external measures did not show any improvements in code quality after the refactoring treatment. However, from internal measures, maintainability index indicated an improvement in code quality of refactored code than non-refactored code and other internal measures did not indicate any positive effect on refactored code.*

## 1. INTRODUCTION

Any useful software system requires constant evolution and changes to meet the ever-changing user needs in a real-world environment. Therefore, the intrinsic property of software system is its need to evolve. As the software system is enhanced, modified and adapted to new requirements, the code become more complex and drifts away from its original design. Because of this, the major part of total software development cost is devoted to software maintenance. Maintenance of software is reported as a serious cost factor [1] and as stated in [2], over 90% of the software development cost is for software maintenance.

While a software system is evolving, maintaining the software quality is one of the vital factors in software maintenance process. The reason is, quality software are robust, reliable and easy to maintain, and therefore, reduce the cost of software maintenance [3]. Software quality can be described as the conformance to functional requirements and non-functional requirements, which are related to characteristics described in the ISO-9126 standard namely reliability, usability, efficiency, maintainability and portability [4]. In addition, factors that affect software quality can be classified into two groups [5]: factors that can be directly measured i.e. internal quality attributes (e.g. Coupling, Cohesion, LOC and etc.) and factors that can be measured only indirectly i.e. external quality attributes (e.g. understandability, analyzability and etc.).

Software maintenance best practices are arising with the purpose of a better evolution of software while preserving the quality of software systems. Thus, the one solution proposed to reduce the software maintenance effort while maintaining the software quality is software refactoring (Fowler, 2000), which is a method of continuous restructure of code according to implicit micro design rules. According to the Fowler's definition (Fowler, 2000), refactoring is the change made to the internal structure of the software system by removing bad smells or problematic places in the source code to make it easier to understand and cheaper to modify without changing its observable behavior.

Although, the refactoring is by definition supposed to improve the maintainability of a software product, its effect on other quality aspects is unclear. Therefore, there are hot and controversial issues in refactoring. As stated by Mens and Tourwé [1], refactoring is assumed to be positively affect non-functional aspects, like extensibility, modularity, reusability, complexity, maintainability, and efficiency. Bios and Mens (2003) performed a return on investment analysis in an open source project, in order to estimate savings in effort, given a specific code change. They found that most of the time, refactoring has beneficial impacts on maintenance activities, and thus are motivated from an economical perspective. However, additional negative aspects of refactoring are reported too [1]. They consist of additional memory consumption, higher power consumption, longer execution time, and lower suitability for safety critical applications.

Several studies have been conducted to evaluate the impact of refactoring of software quality ([7]; [8]). Even though those studies claim that refactoring improves the quality of software, most of them did not provide any quantitative evidence. Therefore, the empirical evidence of the effect of refactoring is rare to be found [9]. As mentioned in [10] 'effect of a refactoring on the software quality' is one of the open issues that remain to be solved.

Altogether, the real advantages of refactoring are still to be fully assessed. Regarding the quality, it appears to be a convergence of positive remarks, still, without solid quantification. In addition there are few quantitative evaluations of impact of each refactoring techniques to the software quality. It is sometimes difficult to judge whether the refactoring in question should be applied or not without knowing the effect accurately. Especially in software development industry, from the viewpoint of project managers, it is imperative to evaluate quantitatively the effect of refactoring on program before applying it. Without knowing which refactoring technique will be more beneficial in terms of quality, managers cannot judge whether they should go for refactoring or not because they have to be cost sensitive. Therefore, there is a need of a study which can evaluate quantitatively the impact of each refactoring technique on quality of code. The objective of this study is to evaluate the effect of refactoring on code quality improvement in order to decide whether the cost and the time put into refactoring are worthwhile.

The reminder of this paper structured as follows: Section 2 provides a summary of relevant literature along with comprehensive review of relevant work. Research methodology used for the research is described in Section 3. Section 4 provides experimental design for the evaluation of impact of refactoring using external measures and the measurement procedure used for the analysis of the impact of refactoring using internal measures. Analysis of research findings is presented in section 5 and 6 for external measures and internal measures respectively. Finally, the section 7 provides the discussion of results and section 8 provides the conclusions and suggestions for future research that can be pursued in this area.

## 2. RELATED WORK

A growing number of studies address the relationship between refactoring and the internal structure of source code and its impact on program understanding, software quality, and the evolution of a software design.

Studies which have evaluated the impact of refactoring on software quality can be categorized into three main categories according to the focused quality factors: internal quality factors, external quality factors and combination of both quality factors.

Even though some of those studies claim that refactoring improves the quality of software, most of them did not provide quantitative evidence. Few researches quantitatively evaluated whether refactoring indeed improves quality (ex. [7]; [8]).

Among them, significant number of studies evaluated quantitatively the impact of refactoring using internal quality attributes. Bois and Mens [6] proposed a technique using metrics to analyse the refactoring impact on internal quality metrics as indicators of quality factors. They proposed formalism based on abstract syntax tree representation of the source-code, extended with cross-references to describe the impact of refactoring on internal program quality. They focused on three refactoring methods: "Encapsulate Filed", "Pull up Method" and "Extract Method". However, they did not provide any experimental validation. Finally, the results of their work showed both positive and negative impacts on the studied measures. Stroggylos and Spinellis [10] analyzed source code version control system logs of four popular open source software systems to detect changes marked as refactoring and examine their effects on software metrics. They finally came up with a conclusion that refactoring does not improve quality of a system in a measurable way. Bois et al. [11] developed practical guidelines in order to applying to refactoring methods to improve coupling and cohesion characteristics and validated these guidelines on an open source software system. There were only five refactoring techniques under study: Extract Method, Move Method, Replace Method with Method Object, Replace Data Value with Object, and Extract Class. They assumed that coupling and cohesion are internal quality attributes which are generally recognized as indicators for software maintainability.  At the end they came up with results that the effect of refactoring on coupling and cohesion measures can be ranged from negative to positive. Kannangara and Wijayanayake [12] evaluated both overall and individual impact of selected refactoring techniques. Ten refactoring techniques were evaluated by them through experiments and assessed five internal measures: Maintainability Index, Cyclomatic Complexity, Depth of Inheritance, Class Coupling and Lines of Code. They used source codes developed using C#.net and internal measures were extracted through Visual Studio IDE. According to their findings, only maintainability index indicated an improvement in code quality of refactored code than non-refactored code and other internal measures did not indicate any positive effect on refactored code.

Few other studies took the approach of assessing the refactoring effects on external software quality attributes. Geppert et al. [13] empirically investigated the impact of refactoring on changeability. This study found that the customer reported defect rates and change effort decreased in the post-refactoring releases. The effect of refactoring on maintainability and modifiability was investigated by Wilking et al. [8] through an empirical evaluation.

Maintainability was tested by randomly inserted defects into the code and measuring the time needed to fix them. Modifiability was tested by adding new requirements and measuring the time and Line of Code (LOC) metric needed to implement them. Their findings on maintainability test show a slight advantage for refactoring and Modifiability test shows disadvantage for refactoring. Kannangara and Wijayanayake [14] evaluated ten refactoring techniques separately using four external quality factors. Outcome of their study indicates that there is no quality improvement in most of the refactoring techniques that they have tested.

Other remaining studies used the approach of assessing the impact of refactoring on internal attributes as indicators of external software attributes. To do so, they defined and relied on relationships between internal and external attributes defined by different authors (ex. [16]). Kataoka et al. [7] proposed coupling metrics as a quantitative evaluation method to measure the effect of refactoring on program maintainability. For the purpose of validation they analyzed a C++ program for two refactoring techniques: Extract Method and Extract Class which developed by a single developer, however did not provide any information on the development environment. Thus, it is questionable if their findings are valid in a different context where development teams follow a structured process and use common software engineering practices for knowledge sharing. Moser et al. [17] proposed a methodology to assess whether the refactoring improves reusability and promotes ad-hoc reuse in an Extreme Programming (XP)-like development environment. They focused on internal software metrics that are considered to be relevant to reusability based on metric interpretation of Dandashi and Rine's (2002) work. They came up with a conclusion that refactoring has a positive effect on reusability. The impact of refactoring on development productivity and internal code quality attributes was analyzed by Moser et al. [17]. A case study has been conducted to assess the impact of refactoring in a close-to industrial environment and the collected measures were Effort (hour), and Productivity (LOC). Results indicate that refactoring not only increases aspects of software quality, but also improves productivity. Alshayeb [3] quantitatively assessed the effect of refactoring on different external quality attributes: Adaptability, Maintainability, Understandability, Reusability, and Testability using software matrices based on metric interpretation of [16]. However, this study didn't prove that refactoring improves external quality of the software. Shatnawi and Li [18] studied the effect of software refactoring on software quality. They have conducted the study on a larger number of refactoring techniques (43 refactoring) and measured four external quality factors indirectly using nine different internal software quality measures based on Quality Model for Object Oriented Design (QMOOD). They had provided details of findings as heuristics that can help software developers make to more informed decisions about what refactoring techniques to perform in regard to improve a particular quality factor. They have validated the proposed heuristics in an empirical setting on two open-source systems. They found that the majority of refactoring heuristics do improve the quality; however some heuristics do not have a positive impact on all software quality factors.

Several concerns in those studies are:

- All of these previous studies did not come up with the same conclusions regarding the impact of refactoring. Therefore, there is a need of analyzing the impact of refactoring further.
- Most of the studies which evaluated external quality factors did it by using internal quality factors and majority of them have used quality models. Therefore, their research findings totally depend on the validity of those quality models.
- Those who evaluated external quality factors only focused on one or two external quality factors. None of them have focused on ISO quality factors or other accepted quality model for selecting quality factors.

- Most of them didn't measure both internal and external measures separately in their studies.
- Except one study [18] all the other studies used less number of refactoring techniques for their evaluation. Most of them did not provide any valid justification when selecting refactoring techniques for their study.

To overcome above issues, this study was designed using ten refactoring techniques and focused on both external and internal quality factors.

# 3. METHODOLOGY

After reviewing relevant literature, the main objectives of this study was defined to quantitatively assess the effect of refactoring on code quality using different external and internal measures separately in order to decide whether the cost and the time put into refactoring are worthwhile.

To achieve above objective, study was carried out separately using two measurements: external measurements and internal measurements. Selection of both measures is mostly influenced by previous related studies. Most of previous studies either measured internal or external measures and some of them interpreted external measures by using internal measures. Therefore, this study mainly focuses on measuring both measures separately in order to assess the impact of refactoring on code quality. As external measurements, external quality factors have been used and as internal measurements, code metrics have been used.

Furthermore, quantitative research approach was selected for this study. As the experiential evidence of the effect of refactoring is rarer to be found and those experiments were ended up with mixed picture of refactoring, experimental research approach is selected to quantitatively access the impact of refactoring on code quality. Only one previous study [8] was used experimental research approach to evaluate the impact of refactoring on quality factors. They were not able to prove refactoring improves code quality. Hence, that become a good reason for the selection of experimental research approach.

When measuring both external and internal measures, same refactored and non-refactored source codes were used and the outcomes were analysed.

## 3.1. Selected Refactoring Techniques

Fowler (2000) proposed 72 refactoring techniques in his catalogue of refactoring. Among the studies which have evaluated the impact of refactoring, the most recent study [18] present evaluation of 43 refactoring techniques among 72 refactoring techniques in Fowler's (2000) catalogue. Evaluated refactoring techniques were ranked according to the impact of code quality. Therefore, for this study, ten refactoring techniques were selected from Shatnawi and Li's [18] study which were ranked as having a high impact.

Selected Refactoring Techniques are:

R1- Introduce Local Extension
R2- Duplicate Observed Data
R3- Replace Type Code with Subclasses
R4- Replace Type Code with State/Strategy
R5- Replace Conditional with Polymorphism

R6- Introduce Null Object
R7- Extract Subclass
R8- Extract Interface
R9- Form Template Method
R10- Push Down Method

## 3.2.  Selection of Source Code Development Environment and Source Code

Refactoring is a technique which is mainly related to object oriented programming. Therefore, the selection of development environment and programming language was done mainly based on the above reason.

Java, C# and C++ are some of the most popular object oriented programming languages which are being used in the current IT industry. Among those, Java and C++ are the commonly used programming languages in previous studies which evaluated the impact of refactoring on code quality improvement ([7]; [18]).

Therefore, C# was selected as the programming language and Visual Studio as the development tool for this study.

In order to apply 10 refactoring techniques a small scale project with bad smells was selected as the source code. The selected application was a system which was implemented in the Department of Industrial Management, University of Kelaniya and used by academic staff at the department to schedule their personal and professional events and to manage their online documents repository. The source code contained around 4500 lines of codes. The relevant bad smells were identified and all the selected refactoring techniques were applied to the source code.

## 3.3. Selected External Measures

Most of the previous studies claimed that refactoring improves the software quality. Software quality is a general term and it can define with several quality attributes. Thus, all of those arguments should be valid with any software quality attribute.

This study was designed to validate those arguments and several software quality attributes were selected from ISO quality model [4]. The reason for selecting ISO quality model was that as stated in [19], it is the most useful one since it has been built based on an international consensus and agreement from all the country members of the ISO organization. Following are the external quality attributes which were used for this study:

(1) Maintainability:   A set of attributes that bears the effort needed to make specified modifications [4]. Following sub characteristics were tested in this study.
    a.   Analysability
    b.   Changeability
(2) Efficiency:  Efficiency is a set of attributes that bear on the relationship between the level of performance of the software and the number of resources used, under stated conditions [4]. Following sub characteristic were tested in this study.
    a.   Resource Utilization
    b.   Time behaviour

## 3.4. Selected Internal Measures

Code metrics have been selected as internal measures to judge the impact of refactoring on code quality. As this study is strived to measure the maintainability of software, metrics which can measure maintainability and complexity of code is considered as main selection criteria for the selection of internal measures. Therefore, the selected code metrics were [23],

(1) Maintainability Index
(2) Cyclomatic Complexity
(3) Depth of Inheritance
(4) Class Coupling
(5) Line of Code

It can be validated/invalidated the assumption that refactoring improves code quality by comparing values which obtained from above metrics.

# 4. EXPERIMENTAL DESIGN

## 4.1. Design of Experiment for External Measures

An experiment was carried out to evaluate the impact of refactoring using external measures. The experiment consist of a group of participants with the same application developed using C#.net. One group was assigned refactored code with selected refactoring technique/s while the rest was assigned source code without refactoring. The assignment to a treatment and control groups were done at random.

As subjects was randomly assigned to two groups and one group received an experimental treatment while the other group(the control group) received no treatment, experimental design type was selected as two group post-test-only randomized experiment.

### 4.1.1. Sample Selection

The major skill required with participants was their programming skill. Therefore, the selection criterion of target population was programming skills. Current undergraduates and recently graduated students of University of Kelaniya were selected as the population for experimental sample selection.

The sample selection procedure was carried out based on two criteria. Those are,

• Based on semester examination results for programming related subjects
• Based on survey done in order to identify student's familiarity of C#.Net and Object Oriented Concepts: Online questionnaire was designed to gather responses.

After collecting both data, students' results and responses were scaled to ten. Average values for each student was calculated and ranked them based on average value. Then the selection of students for the experiment was done according to their rank starting from top ranks.

For the experiment or to analyse all the selected refactoring techniques together, size of the group was decided as 10 members per one group. Due to availability of limited resources at Undergraduate laboratories and controlling of large groups is not possible with available human resources, group number was limited to 10.

### 4.1.2. Variables and Measurements

Independent Variables:

The independent variable for this experiment is the treatment which is a single, dichotomous factor. Either a participant is assigned to a group which uses a refactored code or to a group which uses a code without refactoring, in order to rule out the placebo effect which is known as a phenomenon which may result in some therapeutic effect if subjects is given control [23].

Dependent Variables:

Dependent variables for this experiment are,

- Marks obtained for question paper
- Time need to fix bugs
- Execution Time
- Memory Consumption

### 4.1.3. Research Hypothesis

This study was aimed at presenting evidence that would allow rejecting (or accepting) the following four hypotheses:

Analysability
$H_0A$: Analysability of refactored code is lower than un-refactored code.
$H_1A$: Analysability of refactored code is higher than un-refactored code.
Changeability
$H_0B$: Changeability of refactored code is difficult than un-refactored code.
$H_1B$: Changeability of refactored code is easier than un-refactored code.
Time Behaviour
$H_0C$: Response time of refactored code is longer than un-refactored code.
$H_1C$: Response time of refactored code is shorter than un-refactored code.
Resource Utilization
$H_0D$: Efficient utilization of computer Resources is lower for refactored code than un-refactored code.
$H_1D$: Efficient utilization of computer Resources is higher r for refactored code than un-refactored code.

### 4.1.4. General Procedure

The first step of each experiment was done with controlled and experimental groups. The second step for each experiment was carried out in a software testing environment, to collect resource utilization and time behaviour measures.

Step 1:

The execution of the experiment was started with an oral presentation by introducing application which is being used for the experiment, the experimental environment with procedure, and the general conditions of the experiment.

After that, an initial test was carried out to assess the impact of refactoring on code analysability. Initially five minutes were provided to both groups to be familiar with source code. One group was a control group which was assigned to un-refactored code and the other group was an experimental group which was assigned to a refactored code. After that a question paper which contained multiple choice questions and short answer questions related to source code was distributed to participants and 30 minutes were provided to answer questions by referring the code. At the end of the experiment, question papers were evaluated and marks were recorded for the analysis.

To analyse the impact of refactoring on changeability next experiment was carried out. Source codes with randomly inserted bugs were provided to both experimental and controlled groups. Error descriptions were provided for semantic errors. Participants were worked on fixing bugs and 90 minutes of time frame was provided. Time used to fix bugs was recorded for analysis.

Step 2:

In order to measure resource utilization; memory consumption of software application to execute one selected part from the application was measured and to measure time behaviour, the selected part execution time was measured [4]. When selecting a part from the application, a piece of code which is mostly affected by applied refactoring techniques was selected. Programs were modified to execute automatically 1000 times to collect accurate figures related to execution time and memory consumption during the selected task execution.

### 4.2. Design of experiment for internal measures

As mentioned earlier Visual Studio was selected as the extraction tool for internal measures. To measure the impact of refactoring the same original and refactored source codes which used for previous experiment were used. Internal measures were generated for both source codes and recorded for further analysis.

## 5. ANALYSIS OF DATA – EXTERNAL MEASURES

As the research is quantitative and involves ratio data, parametric statistical test was used for hypothesis testing.

### 5.1. Data Analysis for Analysability

Analysability was measured by using marks obtained by each group member for the given question paper. Same question paper which contained 15 multiple choice and short answer questions was distributed to both controlled and experimental groups. The time duration for question paper was 30 minutes and 1 mark was given to each correct answer. For short answer questions also there was only single correct answer. Therefore, if that answer was there 1 mark was given and otherwise 0 marks. Hypothesis which was tested for Analyzability is that "analysability of refactored code is higher than non-refactored code". Table 1 summarized results of hypothesis testing.

Table 1 Hypothesis test results for Analyzability

| Level of Significance | 0.05 |
|---|---|
| Controlled Group | |
| Sample Size | 10 |
| Sample Mean | 7.1 |
| Sample Standard Deviation | 3.6 |
| Experimental Group | |
| Sample Size | 9 |
| Sample Mean | 6.63 |
| Sample Standard Deviation | 2.13 |
| $t$ Test Statistic | 0.344524 |
| $p$-Value | 0.466775 |
| Do not reject the null hypothesis | |

The assumption of better analysability cannot be answered according to hypothesis test results; because there is insufficient statistical evidence to claim marks obtained by experimental group is higher than control group. In fact it is lesser in experimental group. Therefore it can be stated that refactoring does not significantly affect analysability of code of small scale system.

## 5.2. Data Analysis for Changeability

The measurement of changeability, which consisted of a random insertion of two non-syntactical errors and one new requirement was measured by using time needed to fix bugs in minutes. Hypothesis which was tested under Changeability is that "changeability of refactored code is easier than non-refactored code". Table 2 summarized results of hypothesis testing.

Table 2 Hypothesis Test Results for Changeability

| Level of Significance | 0.05 |
|---|---|
| Controlled Group | |
| Sample Size | 10 |
| Sample Mean | 59 |
| Sample Standard Deviation | 26.27 |
| Experimental Group | |
| Sample Size | 10 |
| Sample Mean | 77 |
| Sample Standard Deviation | 27.72 |
| $t$ Test Statistic | -1.57325 |
| $p$-Value | 0.933464 |
| Do not reject the null hypothesis | |

The assumption of better changeability thus cannot be answered according to hypothesis testing; because, there is insufficient statistical evidence to claim a time spent by experimental group is less than control group. Therefore, it can be stated that refactoring does not significantly affect changeability of code of small scale application.

### 5.3. Data Analysis of Time Behaviour

The measurement of time behaviour was measured by recording task execution time. Piece of code which is highly affected by refactoring treatment was selected and the task which is related to that code segment was selected for testing. Both pre and post refactored programs were modified to execute 1000 times automatically. Results were recorded in milliseconds. Outliers were detected from 1000 sample size from both samples. A hypothesis which was tested for Time Behaviour is that "response time of refactored code is less than non-refactored code". Table 3 summarized results of hypothesis testing.

Table 3 Hypothesis Test Results for Time Behavior

| Level of Significance | 0.05 |
|---|---|
| Original Code | |
| Sample Size | 994 |
| Sample Mean | 61.18 |
| Population Standard Deviation | 21.22 |
| Refactored Code | |
| Sample Size | 985 |
| Sample Mean | 75.71 |
| Population Standard Deviation | 20 |
| Z-Test Statistic | -15.7109 |
| $p$-Value | 1 |
| Do not reject the null hypothesis | |

The assumption of better time behaviour of refactored code thus cannot be answered according to hypothesis testing; because, there is insufficient statistical evidence to claim that task execution time for refactored code is less than code without refactoring. Therefore, the conclusion of better time behaviour is not facilitated by refactoring for small scale applications.

### 5.4. Data Analysis for Resource Utilization

Resource utilization was measured by using memory consumption of program while it is executing. Piece of code which is highly affected by refactoring treatment was selected and the task which is related to that code segment was selected for testing. Both pre and post refactored programs were changed to execute 1000 time automatically. Results were recorded in bytes. Outliers were detected from 1000 sample size from both samples. A hypothesis which was tested for Resource Utilization is "efficient utilization of computer Resources is higher for refactored code than non-refactored code". Table 4 summarized results of hypothesis testing.

Table 4 Hypotheses Testing results for Resource Utilization – Memory Consumption

| Level of Significance | 0.05 |
|---|---|
| Original Code | |
| Sample Size | 1000 |
| Sample Mean | 370970.4 |
| Population Standard Deviation | 159046.9 |
| Refactored Code | |
| Sample Size | 1000 |
| Sample Mean | 377310.3 |
| Population Standard Deviation | 162510.2 |
| Z-Test Statistic | -0.88169 |
| *p*-Value | 0.811027 |
| Do not reject the null hypothesis | |

The assumption of better resource utilization of refactored code thus cannot be answered according to hypothesis testing; because according to the hypothesis test results, there is insufficient statistical evidence to claim a minimum memory allocation for refactored code than code without refactoring. Therefore, the conclusion of better resource utilization is not facilitated by refactoring in small scale systems.

## 5.5. Summary of Results

Table 5 shows the summary of hypothesis testing results of the impact of refactoring on code quality measured by using external measures. In the table symbols are represented as follows.

- Improvement:    '+'
- Deteriorate:    '-'
- No impact:    '0'

Table 5 Summary of the effect of refactoring on code quality using external measures

| Analysability | Changeability | Time Behaviour | Resource Utilization | No Improvements | Improvements |
|---|---|---|---|---|---|
| - | - | - | - | 4 | 0 |

Here it can be noticed that none of the external measures show improvements in code quality when all the selected refactoring techniques are applied together.

## 6. ANALYSIS OF DATA – INTERNAL MEASURES

Code metrics values were generated from both refactored and non-refactored codes and Table 6 summarized the results with percentage of improvement in code metrics after refactoring code using all the selected refactoring techniques. When generating results following are the methods/formulas used by the tool to give the code metrics values [20].

- Maintainability Index = MAX(0,(171 - 5.2 * ln(Halstead Volume) - 0.23 * (Cyclomatic Complexity) - 16.2 * ln(Lines of Code))*100 / 171)

- Class Coupling: Measures the coupling to unique classes through parameters, local variables, return types, method calls, generic or template instantiations, base classes, interface implementations, fields defined on external types, and attribute decoration.
- Cyclomatic Complexity: Measures the structural complexity of the code by calculating the number of different code paths in the flow of the program
- Depth of Inheritance: Indicates the number of class definitions that extend to the root of the class hierarchy.
- Lines of Code – Indicates the approximate number of lines in the code

Table 6 Code Metrics Values for the all the refactoring techniques

| Internal Measure | Non-refactored Code | Refactored Code | % Change |
|---|---|---|---|
| Maintainability Index | 69 | 72 | 4% |
| Cyclomatic Complexity | 1367 | 1436 | -5% |
| Depth of Inheritance | 5 | 5 | 0% |
| Class Coupling | 221 | 237 | -7% |
| Lines of Code | 4922 | 5005 | -2% |

The higher value for maintainability index (MI) indicates the higher maintainability of the code. Therefore, as MI for refactored code is higher than non-refactored code, it can be concluded that refactored codes are maintainable compared to non-refactored code. Cyclomatic complexity, class coupling and line of code metrics are preferred to have lower values as metrics values for highly maintainable source codes. However, for refactored codes, values for these metrics are higher than non-refactored code. Therefore, the refactored code's maintainability is relatively lower than non-refactored code when considering cyclomatic complexity, class coupling and line of code metrics.

Furthermore, high number of deteriorates of quality than the quality improvements can be noticed in Table 6 according to the percentage change of code metrics values. Therefore, it can be stated that the impact of refactoring on code quality is not showing positive results on majority of internal measures.

## 7. DISCUSSION AND RESEARCH FINDINGS

### 7.1. Discussion on Impact of Refactoring on Code Quality Using External Measures

Impact of refactoring on code quality improvement using external measures were measured using four sub quality factors defined in ISO 9126 quality model. Hypothesis test results indicates that there is deteriorate of code quality in refactored code than non-refactored code. Table 7 summarized the findings of analysis of the impact of refactoring on code quality.

Table 7 Summary of effect of refactoring on external measures

| External Measure | No Improvement | Improvements |
|---|---|---|
| Analysability | * | |
| Changeability | * | |
| Time Behaviour | * | |
| Resource Utilization | * | |

Therefore, by using overall analysis and analysis of each refactoring technique, it can be concluded that there is no improvement in code analysability, changeability and time behaviour after applying ten refactoring techniques which was used for this study.

Among related previous studies, Wilking et al. [8] which is the only study which analysed external measure similar to this study, did analysis of impact of refactoring on code maintainability, modifiability and memory consumption. Maintainability and modifiability was negatively affected by refactoring treatment according to their findings. Those measures are more similar to external measures: analysability and changeability used in this study. Thus there is similarity in results of this study and their study for those external measures. However, for memory consumption Wilking et al. [8] got a positive result which is different than results got here.

## 7.2. Discussion on Impact of Refactoring on Code Quality Using Internal Measures

The impact of all the selected 10 refactoring techniques on code quality using internal measures were measured using five internal measures namely: maintainability index, Cyclomatic complexity, Depth of Inheritance, Class coupling and Line of Code.

Table 8 summarized the results obtained by comparing values obtained for code metrics from each refactored and un-refactored codes.

Table 8 Summary of effect of refactoring on internal measures

| Internal measures | No Improvement | Improvements |
|---|---|---|
| Maintainability Index | | * |
| Cyclomatic Complexity | * | |
| Depth of Inheritance | * | |
| Class Coupling | * | |
| Lines of Code | * | |

Cyclomatic complexity, LOC Depth of Inheritance and class coupling indicates there is no quality improvement in source code after refactoring. However, the Maintainability Index shows improvement in code maintainability in refactored code.

The MI is a composite number, based on several unrelated metrics for a software system. It is based on the Halstead Volume (HV) metric, the Cyclomatic Complexity (CC) metric, the average number of lines of code per module (LOC), and optionally the percentage of comment lines per module (COM) [20]. Although MI indicate maintainability is higher for refactored code, Cyclomatic complexity and LOC metrics which are components of MI indicate maintainability of refactored code is lower according to the results of this study. As MI is composite number and there are several arguments about accuracy of MI value [22] , to come up with final conclusion only raw metrics values such as complexity, LOC, coupling and depth of inheritance can be considered. Therefore, the final conclusion that can be derived from internal measures is that there is no quality improvement in source code after refactoring.

Shatnawi and Li [18] did analysis with coupling metric in their study to analyse the impact of refactoring. For the same 10 refactoring techniques, their study got 30% of improvements in coupling metrics and 70% of unchanged. Therefore, there is a slight difference in results of their study and this study. Depth of inheritance also measured in several studies and got negative results ([10]; [17]) which is different than results of this study. Two studies by Moser et al. [15]

and Moser et al. [17] measured line of code metrics and got positive result which is also not compatible with the results of this study.

However, Moser et al. [17] got the same results for complexity metrics as this study. Furthermore, several other studies also claim that value of Coupling metric is decreased after refactoring ([10]; [15]; [17]).

# 8. CONCLUSION AND FUTURE WORK

The main objective of this study was to assess the impact of refactoring on code quality in software maintenance. To achieve that, the research was carried out using two approaches separately. Firstly, the impact of refactoring was assessed using external measures namely; analysability, changeability, time behaviour and resource utilization. Then the impact of refactoring was evaluated using internal measures namely; maintainability index, cyclomatic complexity, depth of inheritance, class coupling and lines of code. Experimental research approach was used to measure both measurements and ten selected refactoring techniques were tested.

According to the experimental results all the external quality factors indicate that there is no quality improvement after refactoring treatment to the source code. Values for internal measures were generated from both refactored and un-refactored codes. According to the analysis the values of Cyclomatic complexity, Class coupling and Line of code metrics indicate that refactoring does not improve the code quality. According to the results of both experiments using two types of measurements: internal and external, this study indicate that refactoring does not improve the code quality.

There are several arguments that may come against this study. The one argument that can come against the experiment is that expert developers would constitute a much better evaluation. Their knowledge on better system design might change the experimental results. However as stated by Höst et al. [24] undergraduate students have comparable assessment ability compared to a professional software developer. Other one is, the work is limited to the refactoring descriptions that appear in Fowler's work and may not apply to variations of these refactorings. Furthermore, the low number of participants for the experiment which can directly affect the sample size for the hypothesis testing.

The results of this study indicate that there is further need of addressing the impact of refactoring. Refactoring techniques used in this study were selected from the ranking done by previous study [18]. Therefore, in the future it is better to conduct a study to find refactoring techniques which are commonly used in industry by a survey. Then analyse the impact of those commonly used refactoring techniques will be more advantageous to the software development industry rather than selecting refactoring techniques subjectively. Furthermore, it would be better that the same experimental setup can be executed in industry environment with the industry experts and with the industry level matured source code.

## Authors

**Miss. Sandeepa Harshanganie Kannagara** holds a Bachelor's degree in Management and Information Technology from the University of Kelaniya with First Class honours. In addition as a professional qualification she has successfully completed the British Computer Society (BCS) examinations and she is a Professional Member of BCS. Presently, she is a Lecturer attached to School of Computing, National School of Business Management, Sri 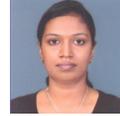 Lanka. Her research findings are published in the International Journal on Advances in ICT for Emerging Regions and presented in several international research conferences. Her research interests are Software Quality, Software Maintenance, Opinion mining and Big Data.

**Dr. W M Janaka I Wijayanayake** received a PhD in Management Information Systems from Tokyo Institute of Technology Japan in 2001. He holds a Bachelor's degree in Industrial Management from the University of Kelaniya, Sri Lanka and Master's degree in Industrial Engineering and Management from Tokyo Institute of Technology. He is currently a Senior Lecture in Information Technology at the department of Industrial Management, 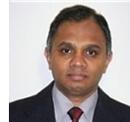 University of Kelaniya Sri Lanka. He has experience in research and teaching in Information Technology related subjects at undergraduate level as well as postgraduate level in Sri Lankan universities and foreign universities. His research findings are published in prestigious journals such as Journal of Information & Management, Journal of DATABASE, Journal of Business Continuity & Emergency Planning, International Journal of Business Continuity and Risk Management and many other journals, and presented in many international academic conferences. His research interests are Data Engineering, Software Engineering, Business Intelligence, Knowledge Management, and Information System Engineering.